\documentclass[twocolumn,a4paper,10pt,showkeys,aps,prc,nofootinbib,superscriptaddress]{revtex4-1}

\usepackage[utf8x]{inputenc}
\usepackage[british]{babel}
\usepackage[british]{isodate}
\usepackage{graphicx}
\usepackage{exscale}
\usepackage{color}
\definecolor{grey}{gray}{0.4}
\usepackage{setspace}
\usepackage{amsmath}
\usepackage{amsfonts}
\usepackage{amssymb}
\usepackage{textcomp}
\usepackage{hyphenat}

\begin{document}
\newcommand{\tgrey}[1]{\textcolor{grey}{#1}}
\newcommand{\tblue}[1]{\textcolor{blue}{#1}}
\newcommand{\tred}[1]{\textcolor{red}{#1}}

\title{Erosion behaviour of composite Al-Cr cathodes in cathodic arc plasmas in inert and reactive atmospheres}

\author{Robert Franz}
\email[Corresponding author: ]{robert.franz@unileoben.ac.at}
\affiliation{Montanuniversit\"{a}t Leoben, Franz-Josef-Strasse 18, 8700 Leoben, Austria}
\author{Francisca Mendez Martin}
\affiliation{Montanuniversit\"{a}t Leoben, Franz-Josef-Strasse 18, 8700 Leoben, Austria}
\author{Gerhard Hawranek}
\affiliation{Montanuniversit\"{a}t Leoben, Franz-Josef-Strasse 18, 8700 Leoben, Austria}
\author{Peter Polcik}
\affiliation{Plansee Composite Materials GmbH, Siebenb\"{u}rgerstrasse 23, 86983 Lechbruck am See, Germany}

\begin{abstract}
Al$_{x}$Cr$_{1-x}$ composite cathodes with Al contents of $x$ = 0.75, 0.5 and 0.25 were exposed to cathodic arc plasmas in Ar, N$_2$ and O$_2$ atmospheres and their erosion behaviour was studied. Cross-sectional analysis of the elemental distribution of the near-surface zone in the cathodes by scanning electron microscopy revealed the formation of a modified layer for all cathodes and atmospheres. Due to intermixing of Al and Cr in the heat-affected zone, intermetallic Al-Cr phases formed as evidenced by X-ray diffraction analysis. Cathode poisoning effects in the reactive N$_2$ and O$_2$ atmospheres were non-uniform as a result of the applied magnetic field configuration. With the exception of oxide islands on Al-rich cathodes, reactive layers were absent in the circular erosion zone, while nitrides and oxides formed in the less eroded centre region of the cathodes. 
\end{abstract}


\cleanlookdateon
\date{\today}

\maketitle

\section{Introduction}

Cathodic arc deposition is a frequently applied physical vapour deposition technique to synthesise hard coating materials like TiAlN \cite{Lugscheider1995,Kimura1999,Mayrhofer2003}, AlCrN \cite{Reiter2005,Reiter2010} or AlCrO \cite{Ramm2007,Ramm2007a,Pohler2014}. In industrial-scale processes, typically alloy or composite cathode materials containing the metal components, i.e.~Ti, Cr or Al for the mentioned examples, are employed as they enable high growth rates and a good reproducibility while the design of the deposition system can be kept relatively simple. The reactive gases N$_2$ or O$_2$ are added during the deposition process to synthesise nitride and/or oxide coating materials. 

The plasma properties present in the cathodic arc plasmas from alloy or composite cathodes have been studied to some extend in order to gain knowledge about the growth conditions encountered in deposition processes utilising cathodic arcs. The materials investigated include Ti-Al \cite{Bilek1998,Zhirkov2014}, Ti-Si \cite{Eriksson2013}, Ti-C \cite{Zhirkov2013} and Al-Cr \cite{Tanaka2015a}. Most of the studies measured the ion charge states or ion energies in vacuum conditions. The influence of an inert or reactive background gas on the cathodic arc plasma from Ti-Al \cite{Zhirkov2015a} and Al-Cr cathodes \cite{Franz2013a,Franz2015} has been evaluated by measuring the ion charge states and energies as a function of the background gas pressure.

From detailed studies using single-element cathodes it is known that the plasma properties are affected by the material properties of the used cathode material and the so-called \textit{cohesive energy rule} was established \cite{Anders2001a}. In the case of alloy or composite cathodes, two or more elements are present in the cathode material and changes in the spatial distribution of the elements and in the structure of the surface-near zone appear due to the frequent heating and cooling of the cathode material that is exposed to the very intense plasma in the cathode spots \cite{Anders2008}. Two decades ago, Eizner \textit{et al.}~noticed that modifications of the surface on eroded Al-Si composite cathodes occurred as compared to the surface of the virgin cathodes \cite{Eizner1996}. The thickness of the modified layer was proportional to the heat conductivity of the cathode material. More recently, the surfaces of a series of cathode materials used for the synthesis of hard coatings has been studied including Al-Cr \cite{Ramm2010,Pohler2011}, Ti-Si \cite{Zhu2010,Zhu2013} and Ti-Al \cite{Rafaja2011,Zhirkov2014} as well as Ti$_3$SiC$_2$ \cite{Zhu2011} and Al-Cr-Si \cite{Paulitsch2014}. The Al-Cr-based systems were exposed to an arc plasma in O$_2$ atmosphere, whereas the Ti-based cathodes were used in N$_2$ containing arc discharges.

The aim of the current work is to study the erosion of composite Al-Cr cathodes with varying composition due to the exposure to arc discharges in inert Ar atmosphere as well as in reactive N$_2$ and O$_2$ atmospheres. The modified surface of the eroded cathodes was analysed by scanning electron microscopy (SEM) and X-ray diffraction (XRD). The results are discussed on the basis of arc cathode erosion characteristics established in literature. Similarities and differences with other material systems and arc discharge conditions are pointed out.

\section{Experimental details}

\begin{figure}
 \centering
 \includegraphics[width=7.8cm]{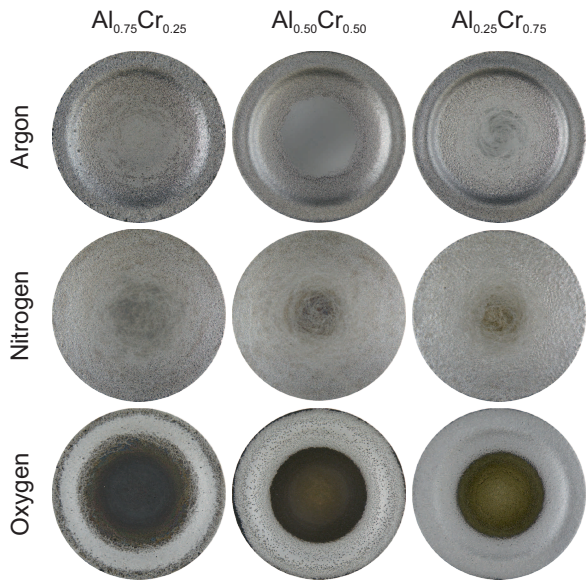}
 \caption{Surface images of the eroded Al$_{x}$Cr$_{1-x}$ arc cathodes vs.~background gas type.}
 \label{fig:AlCr-cathode-erosion_overview}
\end{figure}

The Al$_{x}$Cr$_{1-x}$ cathodes with a diameter of 65 mm were exposed to an arc plasma with a dc current of 50 A. The source (from VTD Vakuumtechnik, Dresden, Germany) was placed in a vacuum chamber with a diameter of 1 m which was pumped to a residual pressure of typically $2.5 \cdot 10^{-4}$ Pa prior to the ignition of the arc discharge. As described in ref.~\cite{Franz2015}, the gas pressure during the experiments was varied between 0.5 and 3.5 Pa. The pressure applied before removing the individual cathodes from the source was 0.5 Pa in the case of Ar used as background gas and 3 Pa in the case of the reactive gases N$_2$ and O$_2$. Images of the surfaces of all cathodes analysed in this work are shown in Fig.~\ref{fig:AlCr-cathode-erosion_overview}. The apparent circular erosion zone is a consequence of the arched magnetic field used in the source to steer the cathode spot motion \cite{Karpov1997,Boxman2005,Anders2008}. The powder-metallurgically prepared Al-Cr composite cathodes had chemical compositions of Al$_{0.75}$Cr$_{0.25}$, Al$_{0.50}$Cr$_{0.50}$ and Al$_{0.25}$Cr$_{0.75}$.

Subsequent to the exposure to the arc plasma, cross-sections of the cathodes were prepared by grinding and polishing to a finish of 1 \textmu m. These cross-sections were analysed by SEM using a Zeiss EVO 50. Backscattered electron (BSE) images were recorded due to the achievable element contrast based on the mass difference between Al and Cr atoms. Elemental distribution maps of Al, Cr, N and O were obtained by energy dispersive X-ray spectroscopy (EDX) using an Oxford Instruments INCA EDX system which was attached to the SEM. In order to identify crystallographic phases present in the near-surface region of the eroded cathodes, a structural analysis by XRD was performed with a D8 Advance diffractometer from Bruker AXS. The device was equipped with a Cu-ceramic tube and a Göbel mirror to enable scans with parallel beam geometry. The intensity of the measurement signal during the $\theta$-2$\theta$ scans was recorded with an energy-dispersive detector (SolX from Bruker AXS).

\section{Results}

\subsection{Cross-sections from erosion zone}

\begin{figure}
 \centering
 \includegraphics[width=7.8cm]{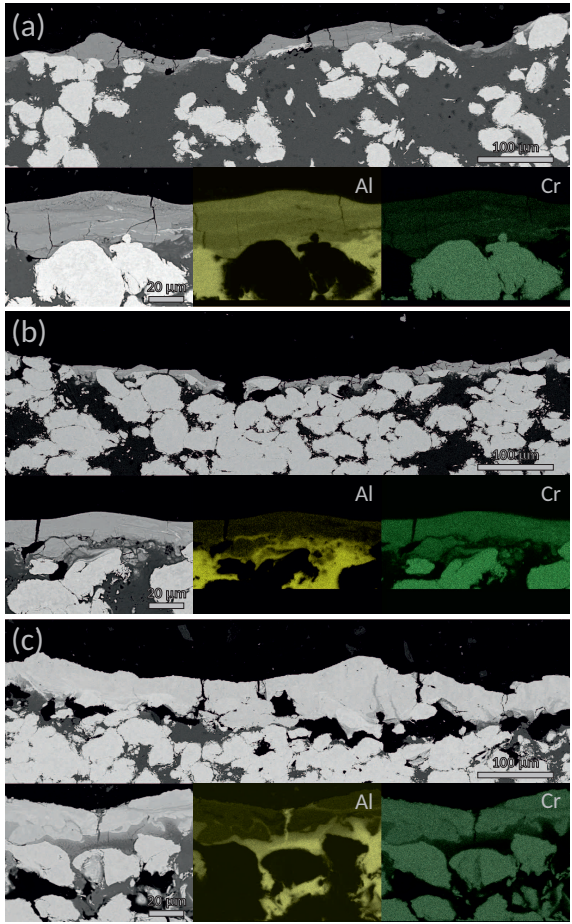}
 \caption{BSE-SEM cross-section images from the erosion zone of the cathodes operated in Ar atmosphere: (a) Al$_{0.75}$Cr$_{0.25}$, (b) Al$_{0.50}$Cr$_{0.50}$ and (c) Al$_{0.25}$Cr$_{0.75}$. The second row of each subfigure shows a surface-near region of the respective cathode with higher magnification and the corresponding elemental distribution maps of Al and Cr.}
 \label{fig:SEM_cross-sections_Ar}
\end{figure}

The composite structure of the virgin cathode material is clearly visible in the cross-section SEM images shown in Fig.~\ref{fig:SEM_cross-sections_Ar} where the Cr grains are embedded in the Al matrix. In the backscattered electron signal Cr-rich regions appear bright, whereas Al-rich areas appear dark. The surface-near region, where erosion of the cathode in the cathodic arc plasma takes place, has a different appearance. Regardless of the cathode composition, the surfaces are largely covered by a layer consisting of both elements as evidenced by the elemental distribution maps of Al and Cr. The cracks visible in the modified layer are part of a crack network as observed in SEM top-view images of the cathode surfaces (not shown). 

The thickness of the modified layer generally varies along all cathode surfaces, but the dimension and the appearance depends on the cathode composition. In the case of the Al-rich Al$_{0.75}$Cr$_{0.25}$ cathode (see Fig.~\ref{fig:SEM_cross-sections_Ar}a), the layer is well adherent to the virgin cathode material and reveals a large variation in thickness with a maximum of about 40 \textmu m. In a few places, the Cr grains are directly exposed to the surface and, hence, to the plasma. The modified layer on the Al$_{0.50}$Cr$_{0.50}$ cathode is generally thinner and less adherent as some pores are visible between the modified layer and the virgin cathode material. The thickest layer of all three cathodes studied was observed on the Al$_{0.25}$Cr$_{0.75}$ cathode with a maximum thickness of up to 100 \textmu m. It is only loosely connected to the underlying virgin cathode material. 

\begin{figure}
 \centering
 \includegraphics[width=7.8cm]{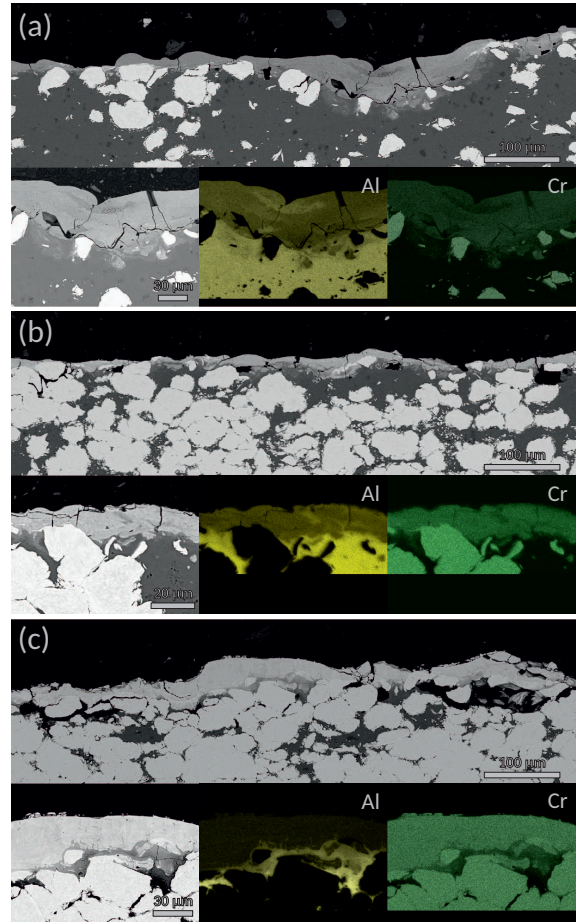}
 \caption{BSE-SEM cross-section images from the erosion zone of the cathodes operated in N$_2$ atmosphere: (a) Al$_{0.75}$Cr$_{0.25}$, (b) Al$_{0.50}$Cr$_{0.50}$ and (c) Al$_{0.25}$Cr$_{0.75}$. The second row of each subfigure shows a surface-near region of the respective cathode with higher magnification and the corresponding elemental distribution maps of Al and Cr.}
 \label{fig:SEM_cross-sections_N2}
\end{figure}

The SEM cross-section images recorded from the AlCr composite cathodes exposed to the arc plasma in N$_2$ atmosphere, as shown in Fig.~\ref{fig:SEM_cross-sections_N2}, are similar to the images from the cathodes eroded in Ar atmosphere. The width of the erosion zone, however, is larger in N$_2$ than in Ar atmosphere (see Fig.~\ref{fig:AlCr-cathode-erosion_overview}). Further, the modified layer near the surface in the case of the Al$_{0.25}$Cr$_{0.75}$ cathode is with a maximum thickness of about 50 \textmu m thinner in N$_2$ than in Ar atmosphere. In terms of chemical composition, the layer again contains a mixture of Al and Cr regardless of the cathode composition as shown in the elemental distribution maps (see Fig.~\ref{fig:SEM_cross-sections_N2}). Even though the cathodes were eroded in a reactive N$_2$ gas atmosphere, no N was detected in the modified layers. 

\begin{figure}[!tb]
 \centering
 \includegraphics[width=7.8cm]{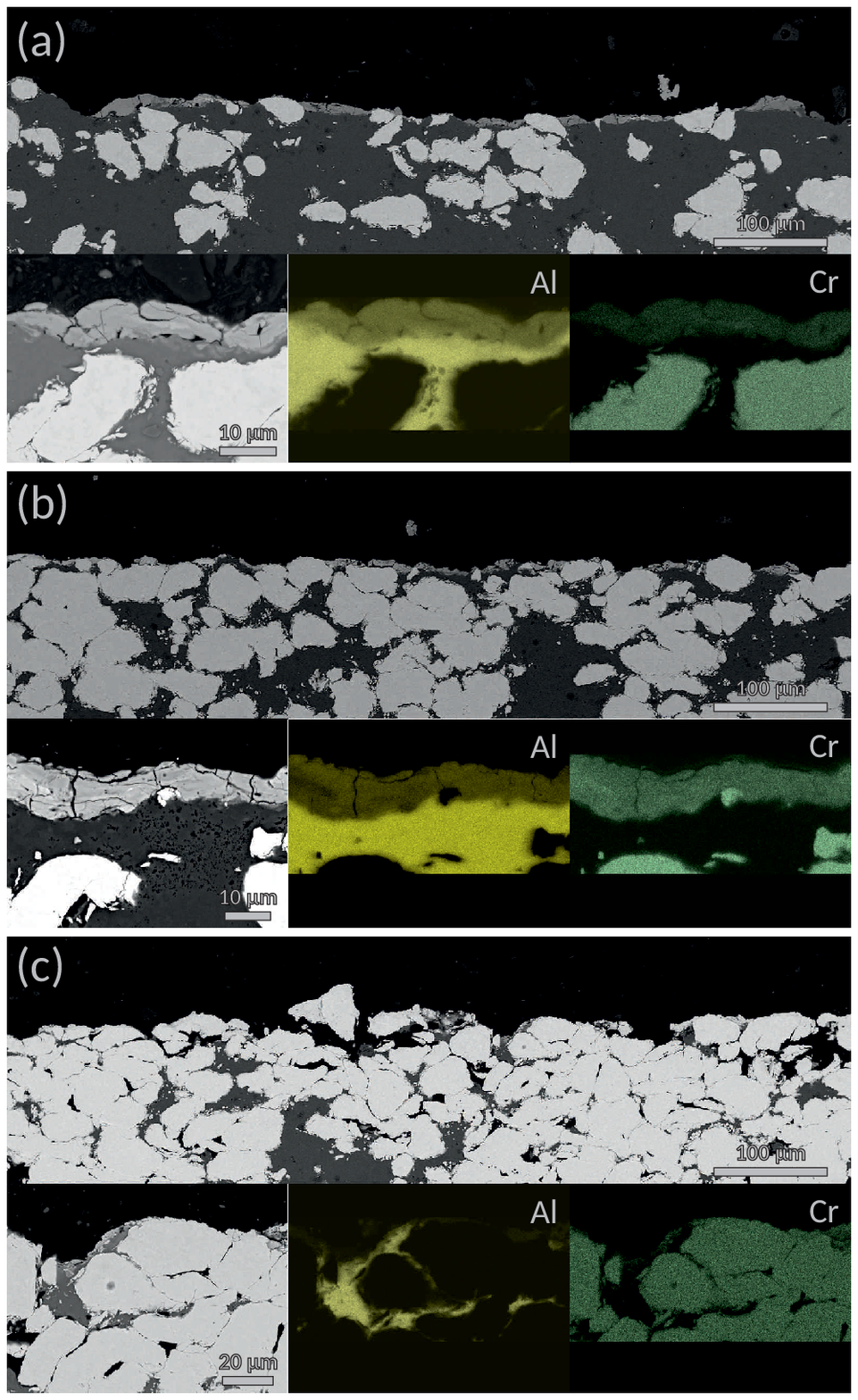}
 \caption{BSE-SEM cross-section images from the erosion zone of the cathodes operated in O$_2$ atmosphere: (a) Al$_{0.75}$Cr$_{0.25}$, (b) Al$_{0.50}$Cr$_{0.50}$ and (c) Al$_{0.25}$Cr$_{0.75}$. The second row of each subfigure shows a surface-near region of the respective cathode with higher magnification and the corresponding elemental distribution maps of Al and Cr.}
 \label{fig:SEM_cross-sections_O2}
\end{figure}

In contrast to the cathodes eroded in Ar and N$_2$ atmospheres, on the surfaces of the cathodes exposed to the arc plasma in O$_2$ atmosphere no or only a thin modified layer was observed as shown in the SEM cross-section images in Fig.~\ref{fig:SEM_cross-sections_O2}. The maximum thickness of the layer is about 20 \textmu m, while it is mainly formed in Al-rich regions. The majority of the Cr grains is exposed directly to the plasma and in the case of the Cr-rich Al$_{0.25}$Cr$_{0.75}$ cathode almost no modified layer was noticed (see Fig.~\ref{fig:SEM_cross-sections_O2}c). 

\begin{figure}
 \centering
 \includegraphics[width=7.8cm]{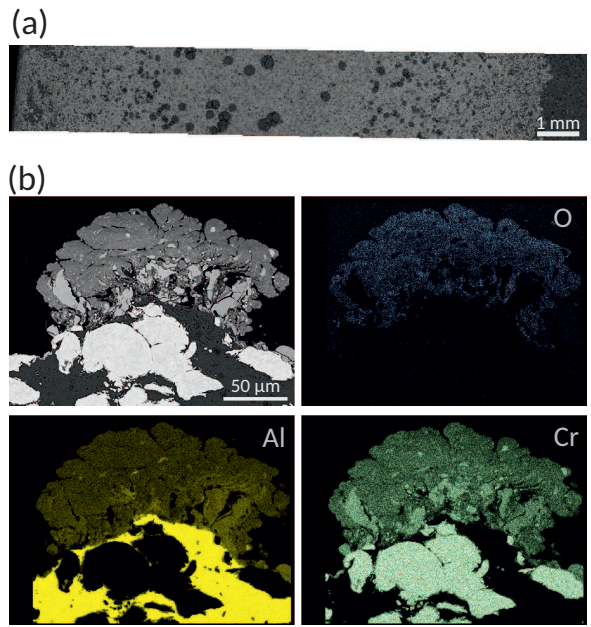}
 \caption{(a) BSE-SEM top-view image of an Al$_{0.50}$Cr$_{0.50}$ cathode after erosion in O$_2$ atmosphere from outer margin (left) to centre (right) and (b) BSE-SEM cross-section image of an oxide island with corresponding elemental distribution maps of O, Al and Cr.}
 \label{fig:SEM_O2_oxide-island}
\end{figure}

Similar to the cathodes in N$_2$ atmosphere, the modified layer contains a mixture of Al and Cr, but generally no O as shown in the elemental distribution maps in Fig.~\ref{fig:SEM_cross-sections_O2} (notice the higher magnification). In the case of the Al$_{0.75}$Cr$_{0.25}$ and Al$_{0.50}$Cr$_{0.50}$ cathodes, black islands formed in the erosion zone as exemplified in the SEM top-view image of Al$_{0.50}$Cr$_{0.50}$ in Fig.~\ref{fig:SEM_O2_oxide-island}a where the entire erosion zone from the outer margin of the cathode to its centre region is displayed. A SEM cross-section of such an island and the corresponding elemental distribution maps are shown in Fig.~\ref{fig:SEM_O2_oxide-island}b. Within the island Al and Cr are largely intermixed, only a few Cr-rich grains remain near the surface of the cathode. O is mainly present in regions of complete intermixture between Al and Cr. These oxide island are the only places where O is encountered in the erosion zone.

\subsection{Cross-sections from cathode centre}

\begin{figure}[!htb]
 \centering
 \includegraphics[width=7.8cm]{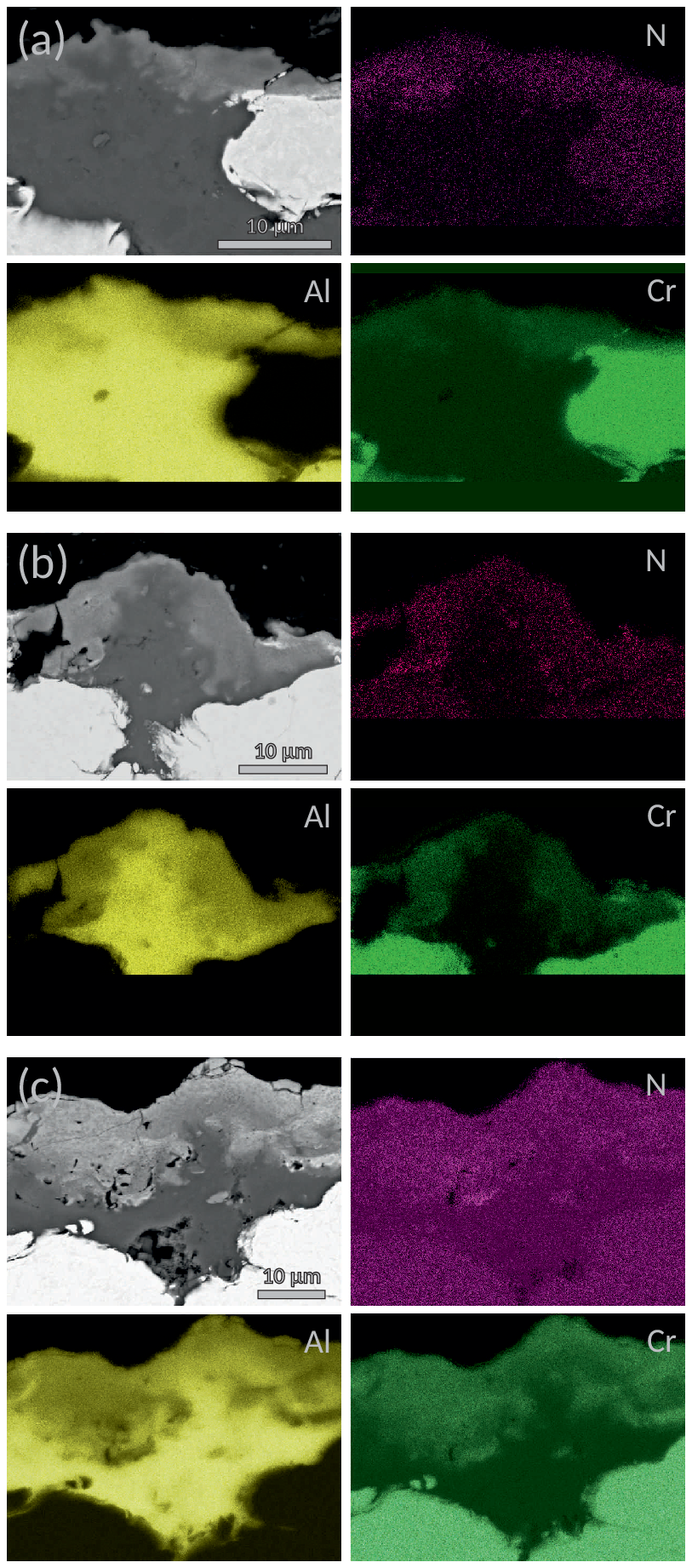}
 \caption{BSE-SEM cross-section images from the centre regions of the cathodes operated in N$_2$ atmosphere: (a) Al$_{0.75}$Cr$_{0.25}$, (b) Al$_{0.50}$Cr$_{0.50}$ and (c) Al$_{0.25}$Cr$_{0.75}$. The corresponding elemental distribution maps of N, Al and Cr for the respective cathode are also shown in each subfigure.}
 \label{fig:SEM_cross-sections_N2_AlCr-centre}
\end{figure}

\begin{figure}[!htb]
 \centering
 \includegraphics[width=7.8cm]{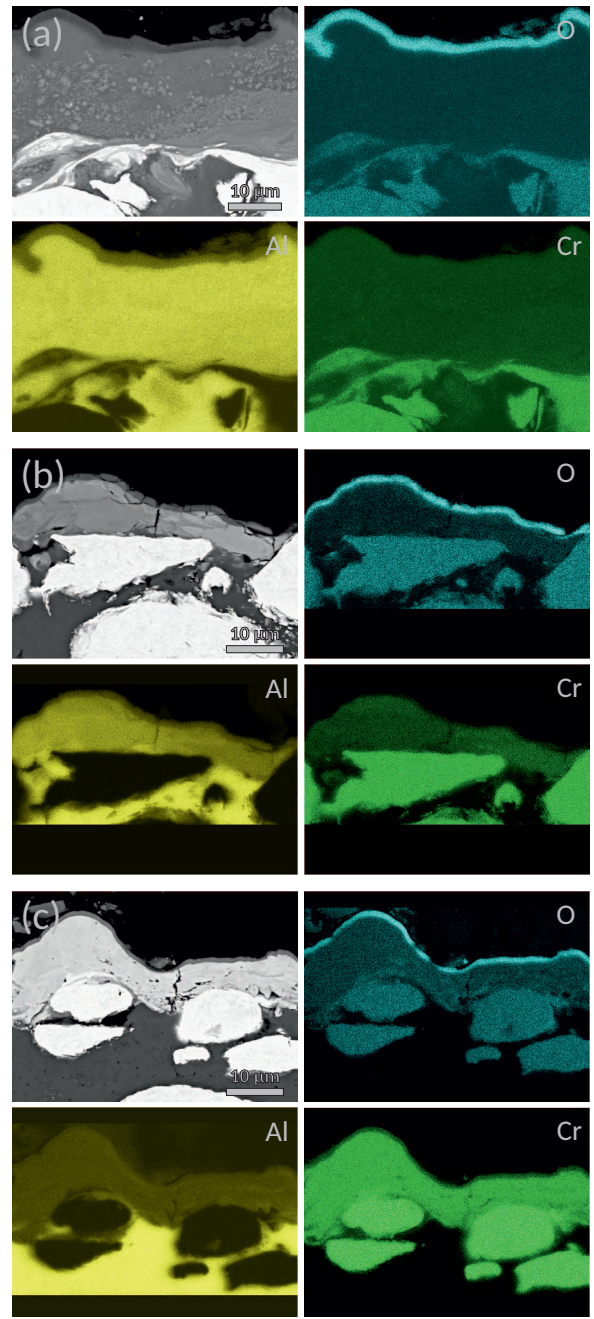}
 \caption{BSE-SEM cross-section images from the centre regions of the cathodes operated in O$_2$ atmosphere: (a) Al$_{0.75}$Cr$_{0.25}$, (b) Al$_{0.50}$Cr$_{0.50}$ and (c) Al$_{0.25}$Cr$_{0.75}$. The corresponding elemental distribution maps of O, Al and Cr for the respective cathode are also shown in each subfigure.}
 \label{fig:SEM_cross-sections_O2_AlCr-centre}
\end{figure}

As shown in Fig.~\ref{fig:AlCr-cathode-erosion_overview}, the appearance of the surface in the cathode centres differs from the appearance in the circular erosion zone described so far. In the case of the Ar atmosphere, the cathode centres remained largely untouched by the arc plasma and present their virgin state. If the reactive gases N$_2$ and, in particular, O$_2$ are used, a darker zone can be observed in the cathode centres. Figs.~\ref{fig:SEM_cross-sections_N2_AlCr-centre} and \ref{fig:SEM_cross-sections_O2_AlCr-centre} show SEM cross-section images from the centre region of the Al-Cr cathodes that were exposed to arc plasmas in N$_2$ and O$_2$ atmosphere, respectively. Similar to the cross-sections from the erosion zone, the intermixing of Al and Cr in the modified surface layer can be noticed in all cases. In addition, the modified layers contain N and O as shown in their elemental distribution maps. In the case of N, the signal intensity is low and near the detection limit of the used EDX system. The distribution of N is rather homogeneous within the modified layer, whereas an oxide layer covering the surface of the cathodes exposed to the O$_2$ atmosphere can be observed regardless of the cathode composition.

\subsection{Phase analysis}

\begin{figure}[!tb]
 \centering
 \includegraphics[width=7.8cm]{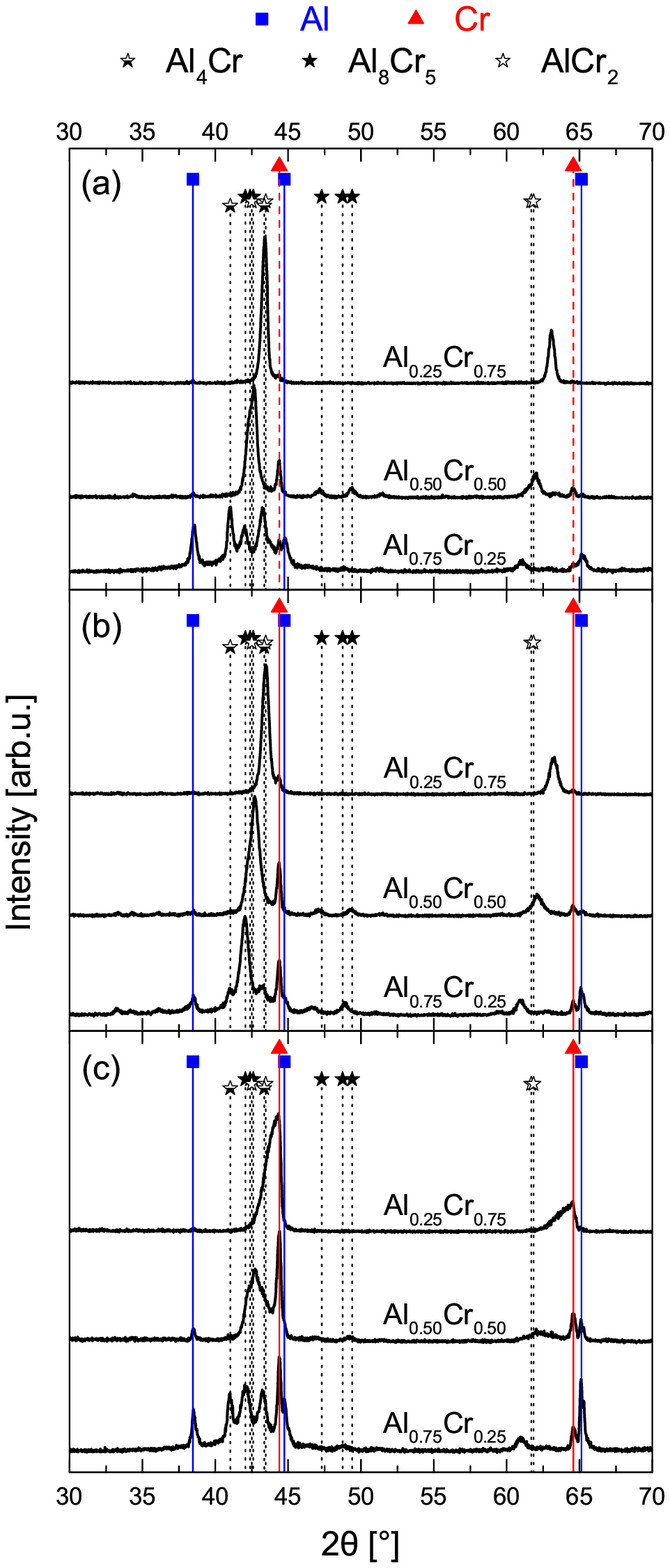}
 \caption{XRD patterns from the erosion zone of the Al$_{x}$Cr$_{1-x}$ cathodes operated in (a) Ar, (b) N$_2$ and (c) O$_2$ atmosphere. Reference peak positions included according to \cite{PDF_AlCr-cathode-erosion}.}
 \label{fig:XRD_erosion-track}
\end{figure}

\begin{figure}[!tb]
 \centering
 \includegraphics[width=7.5cm]{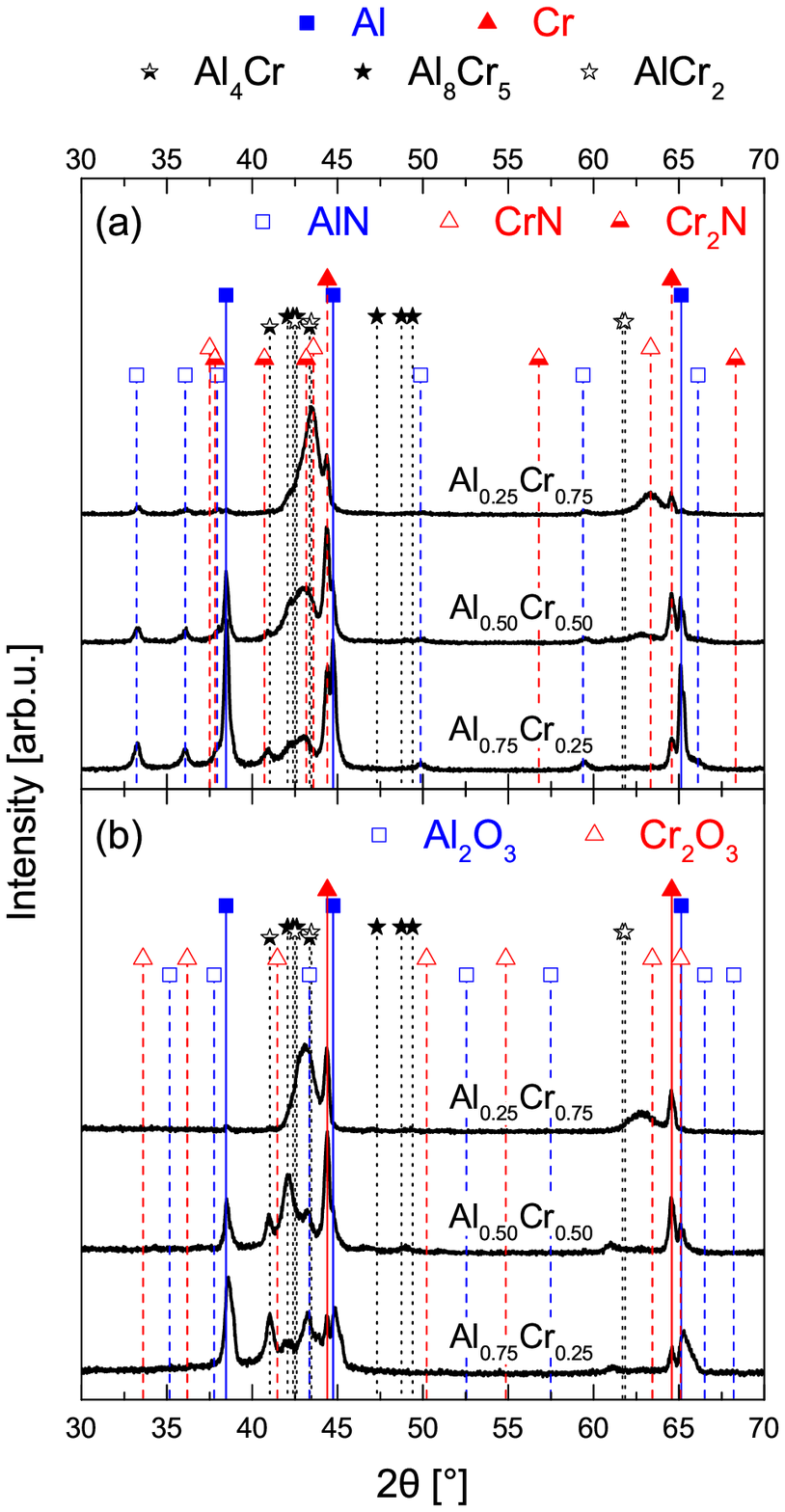}
 \caption{XRD patterns from the centre region of the Al$_{x}$Cr$_{1-x}$ cathodes operated in (a) N$_2$ and (b) O$_2$ atmosphere. Reference peak positions included according to \cite{PDF_AlCr-cathode-erosion}.}
 \label{fig:XRD_cathode-centre}
\end{figure}

Further information can be gained from a phase analysis of the cathode surfaces by XRD as shown for the erosion zones and cathode centres in Figs.~\ref{fig:XRD_erosion-track} and \ref{fig:XRD_cathode-centre}, respectively. In the non-reactive Ar atmosphere, a series of intermetallic phases formed on the surface of the individual cathodes. In the case of the Al$_{0.75}$Cr$_{0.25}$ cathode, Al, Al$_4$Cr, AlCr$_2$ and traces of Cr can be noticed in Fig.~\ref{fig:XRD_erosion-track}a. The peaks of AlCr$_2$ are shifted to smaller angles. No Al phase was present on the surface of Al$_{0.50}$Cr$_{0.50}$, but small amounts of Cr. The main phases, however, were Al$_8$Cr$_5$ and AlCr$_2$. In the case of Al$_{0.25}$Cr$_{0.75}$, mainly a Cr phase with peaks shifted to smaller angles can be noticed which can be understood by the fact that Al was dissolved in the Cr phase.

In N$_2$ atmosphere similar phases on the cathodes as in Ar atmosphere were encountered as shown in Fig.~\ref{fig:XRD_erosion-track}b. On the surface of Al$_{0.75}$Cr$_{0.25}$, less Al and Al$_4$Cr, but more Cr and AlCr$_2$ were present. In addition, Al$_8$Cr$_5$ can be noticed. In the case of Al$_{0.50}$Cr$_{0.50}$, the intensity of the Cr peaks was higher on the cathodes eroded in N$_2$ than in Ar atmosphere. Similarly, small traces of Cr can also be observed in the XRD pattern of Al$_{0.25}$Cr$_{0.75}$.

The phase analysis of the cathodes eroded in O$_2$ also revealed a comparable phase composition. The intensities of the peaks from Al and Cr generally showed a higher intensity for all cathode compositions. The asymmetric peaks recorded in the case of Al$_{0.25}$Cr$_{0.75}$ indicate the presence of two Cr phases, pure Cr and Cr with dissolved Al.

In the centre regions of the cathodes there are less intermetallic Al-Cr phases, but higher fractions of Al and Cr as shown in Fig.~\ref{fig:XRD_cathode-centre} for the cathodes eroded in N$_2$ and O$_2$ atmospheres. On the cathodes exposed to N$_2$, also peaks from AlN can be noticed. Interestingly, there are no indications for crystalline oxide phases like Al$_2$O$_3$ or Cr$_2$O$_3$ on the cathodes exposed to O$_2$ (see Fig.~\ref{fig:XRD_cathode-centre}b). The asymmetry of the Al peaks in the case of Al$_{0.75}$Cr$_{0.25}$ could point towards a cubic Al-O phase in addition to the pure Al phase \cite{Khatibi2011}.

\section{Discussion}

\subsection{Element mixing in erosion zone}

The observed formation of intermetallic Al-Cr phases is in agreement with the phase diagram, e.g.~reported in \cite{Predel2006_Al-Cr}. As the cathode surface is repeatedly heated by the localised plasma in the cathode spots and then cooled down after extinction of the individual spot, the Cr grains dissolve in the surrounding Al matrix. The reaction between Cr and Al and the formation of new phases during dissolving of the Cr grains is governed by diffusion. Since the necessary diffusion length varies with the size of the Cr grains, small grains are fully reacted whereas larger grains only dissolve partially. As a result, not only the Al-Cr phases with similar chemical composition as the average composition of the virgin cathode form, but several intermetallic phases according to the degree of diffusion or intermixing. In the present work, Al$_4$Cr, Al$_8$Cr$_5$ and AlCr$_2$ were identified in the XRD patterns recorded from the erosion zones of the different cathodes. The presence of other intermetallic Al-Cr phases cannot be excluded due to similar structures and, hence, similar peak positions in the XRD patterns. However, it is apparent that the appearance of the cathode surface is different from its virgin state, i.e.~the original state of the cathode is of minor importance for the discharge since the plasma is formed from the modified layer on the cathode surface.

Similar observations were reported in literature for Ti$_{1-x}$Al$_x$ cathodes operated in vacuum conditions where mainly the formation of Ti$_3$Al was noticed \cite{Zhirkov2014}. Other phases could not be identified unambiguously. The opposite effect was observed for Ti$_3$SiC$_2$ cathodes exposed to various discharge conditions \cite{Zhu2011}. In this case, the Ti$_3$SiC$_2$ compound cathode decomposed due to the interaction with the plasma forming a modified layer consisting of TiC$_x$ grains embedded in a matrix of silicide phases.

In general, it can be expected that the diffusion of the elements in the near-surface region of the cathodes is a function of the heat delivered to the cathode. Here, the residence time of the individual cathode spots is decisive since slow moving spots remain longer at a single place and leave larger craters than fast moving spots. With more thermal energy delivered to the cathode surface an enhanced diffusion should take place and a thicker modified layer can form. Applying this relation to the investigated Al-Cr cathodes in this work, the cathode spots in Ar atmosphere have the lowest velocity on the Cr-rich Al$_{0.25}$Cr$_{0.75}$ cathode. But a linear relation with the cathode composition is absent as the observed modified layer on the Al$_{0.50}$Cr$_{0.50}$ is thinner than on the Al-rich Al$_{0.75}$Cr$_{0.25}$ cathode indicating faster moving spots in the case of the former. 

The discussed relation between the thickness of the modified layer and the velocity of the cathode spot motion assumes a constant erosion rate of the cathode regardless of the material encountered there. As it was shown, the appearance of the cathode surface is spatially inhomogeneous and changes with varying cathode composition. Kimblin reported an erosion rate of Al that is three times higher than the erosion rate of Cr \cite{Kimblin1973}. This means that regions in the cathodes without Cr grains will be eroded faster than regions with high concentrations of Cr leading in an uneven erosion profile. However, since the distribution of the Cr grains in the Al matrix is random, the average erosion is independent of the material properties and mainly determined by the spatial distribution of the average plasma density. The latter is mainly influenced by the magnetic field configuration in the present case with a so-called ``steered arc'' \cite{Karpov1997,Boxman2005,Anders2008}.

\subsection{Cathode poisoning}

In contrast to the inert Ar atmosphere or vacuum conditions (not studied in current work), an arc plasma operated in reactive N$_2$ or O$_2$ atmosphere can alter the cathode surface also by reactions between the gas and the solid cathode which is frequently referred to as cathode poisoning \cite{Kim1995,Harris2004,Anders2008}. In the main erosion zones of all Al-Cr cathodes studied here basically no poisoning effects were observed, but only in the centre regions. Due to the applied magnetic field configuration it is less likely that cathode spots are ignited in the centre of the cathode. This results in a reduced erosion from this region and nitrides and oxides were formed as shown in Figs.~\ref{fig:AlCr-cathode-erosion_overview}, \ref{fig:SEM_cross-sections_N2_AlCr-centre} and \ref{fig:SEM_cross-sections_O2_AlCr-centre}. Since the nitrides formed on the AlCr cathodes are conductive, cathode spots are still ignited and erosion from the centre region of the cathodes still takes place to some extent. According to the phase analysis, mainly AlN was present in the modified layer on the cathode surfaces. Cr was reported to react only very littly with N$_2$ in arc discharges \cite{Kuhn1997} which can explain the absence of CrN$_x$ phases.

In contrast, the oxide layer observed is insulating and, hence, prevents the ignition of new cathode spots and the erosion of the cathodes. Underneath the oxide layers shown in Fig.~\ref{fig:SEM_cross-sections_O2_AlCr-centre} there are modified layers noticeable similar to the ones observed on the cathodes exposed to Ar atmosphere. Prior to the O$_2$ pressure reported in the present work, the cathodes were operated at lower O$_2$ pressures as reported in \cite{Franz2015}. Most likely, the oxide layer formed at lower O$_2$ pressures was not stable enough to prevent the ignition of cathode spots completely and, therefore, a modified layer with intermixed Al and Cr was formed.

Within the erosion zone, the appearance of the oxide islands on the Al$_{0.50}$Cr$_{0.50}$ and Al$_{0.75}$Cr$_{0.25}$ cathodes is the only apparent cathode poisoning effect. Such oxide islands have already been reported in literature for Al$_{0.70}$Cr$_{0.30}$ cathodes, but also for other Al-containing cathodes, e.g.~Al$_{0.65}$V$_{0.35}$ \cite{Ramm2010,Pohler2011}. The mechanisms causing the formation of such oxide islands are not well understood. However, the ignition of cathode spots on top of the oxide islands is prevented due to the insulating character of the formed oxide. In the vicinity of these islands, the erosion of the cathode proceeds resulting in the apparent ``growth'' of the islands. With time the neck connecting the island to the bulk of the cathode will be thinned (an early stage is observable in Fig.~\ref{fig:SEM_O2_oxide-island}b) and the oxide island will ultimately be detached due to the progressing erosion. A possibility to suppress the formation of oxide islands is due to the addition of small contents of Si to the cathode as reported in \cite{Paulitsch2014}. Even though the AlCr cathode contained 5 at.\% of Si, it was not incorporated into the deposited coating which was explained by the formation of volatile Si oxides in the discharge.

The formation of intermetallic phases in the erosion zone of the Al-Cr cathodes operated in N$_2$ and O$_2$ atmospheres is comparable to reports in literature investigating Al-Cr-(Si) \cite{Ramm2010,Pohler2011,Paulitsch2014}, Ti-Si \cite{Zhu2010,Zhu2013} or Ti-Al \cite{Rafaja2011} cathodes. The Ti containing cathodes were exposed to arc discharges in N$_2$ atmosphere and the formation of nitrides in the modified layer in addition to intermetallic phases was noticed. This can be understood by the different magnetic field configurations used for the experiments. In the present case, there were no reactive layers in the erosion zone, but in the cathode centre. Apparently, the overall material removal from the Al-Cr cathodes in the erosion zones at the applied N$_2$ and O$_2$ pressures dominates over the formation of reactive layers and mainly metallic surfaces were observed on the Al-Cr cathodes analysed here. However, the presence of the reactive gases N$_2$ and O$_2$ influences the erosion characteristics as the thickness of the modified layers with intermixed Al and Cr is altered in comparison to the modified layers in Ar atmosphere. The slightly lower thickness of the modified layers in N$_2$ than in Ar atmosphere can be explained by the broader erosion zone and the on average shorter residence time of the cathode spots at a single location. In addition, a change from type 1 to type 2 spots should occur with the presence of N$_2$ and (partial) cathode poisoning \cite{Anders2008}. This results in a locally reduced heating of the cathode surface and, hence, less intermixing of Al and Cr. 

In the case of the very thin or absent modified layers on the cathodes exposed to O$_2$ atmosphere, similar effects are active. A superficial oxidation of a few atomic layers up to a few nm in the afterglow of a cathode spot crater reduces the conductivity in this region. Electrons getting trapped subsequently cause a localised enhancement of the electric field which in turn favours the ignition of a new cathode spot plasma at this place and the evaporation of the formed thin oxide layer. As described in \cite{Mustapha2001}, the higher electron emission due to the lower work function of AlO$_x$ than Al \cite{Agarwala1976} is also beneficial. In this way, the cathode spot motion is strongly accelerated within the erosion zone and oxides formed on the heated cathode surface will be removed by the arc discharge. Only at places where a stable oxide of sufficient thickness can form the ignition of cathode spots is prevented due to the insulating nature of the Al-Cr oxides. This means that the erosion of cathode material is stopped in this region and the oxide can remain on the cathode surface as it was the case in the cathode centres and the oxide islands.

\section{Conclusions}

The detailed analysis of composite Al$_{x}$Cr$_{1-x}$ cathodes exposed to arc plasmas revealed the formation of a modified layer on the cathode surfaces, where Al and Cr are intermixed due to the heat transfer from the plasma to the cathode in the cathode spots. The formation of intermetallic Al-Cr phases was noticed for all cathodes and atmospheres tested. The thickest modified layer was observed on the cathodes operated in Ar atmosphere. Due to faster moving cathode spots and, hence, reduced heat transfer in N$_2$ and O$_2$ atmospheres, the thickness of the modified layer was lower in these cases.

As a result of the applied arched magnetic field configuration, non-uniform cathode poisoning effects were observed in the reactive atmospheres. Within the circular main erosion zone no reactive layer formed except oxide islands on the Al-rich cathodes in O$_2$ atmosphere. In the centre region of the cathodes nitrides and oxides were present on the cathodes exposed to arc plasma in N$_2$ and O$_2$ atmosphere, respectively. The continuous, amorphous oxide layer prevented the ignition of new cathode spots due to its insulating character and, hence, the erosion of the cathode in this region.

In general, as the appearance of the cathode surface is modified with respect to its virgin state due to the interaction with the plasma, the modified cathode surface layer determines the arc plasma properties. A detailed knowledge about the phases present on eroded cathodes is therefore of vital importance in order to understand the arc plasma properties.

\section*{Acknowledgements}

The authors wish to thank Sandra Ebner, Matthias Freisinger and Silvia Pölzl for their help with the preparation of the samples. R.~Franz gratefully acknowledges the support of an Erwin Schrödinger Fellowship by the Austrian Science Fund (FWF, Project J3168-N20).


%

\end{document}